\documentclass[twocolumn,twoside,slac_two]{revtex4}
\usepackage{graphicx}
\usepackage{fancyhdr}
\def\etal{et~al.}

\def\snr{SNR~CTA~1}
\def\rxj{RX~J0007.0+7303}
\def\psr{PSR~J0007+7303}
\def\EG{3EG~J0010+7309}

\def\rosat{{\it ROSAT}}
\def\egret{{\it EGRET}}

\newcommand\chandra{{\it Chandra}}
\newcommand\Chandra{{\it Chandra}}

\newcommand\XMM{{\it XMM-Newton}}
\newcommand\veritas{{\it VERITAS}}
\newcommand\VERITAS{{\it VERITAS}}

\newcommand\fermi{{\it Fermi}}

\setcitestyle{numbers}

\begin{document}

\title{VHE Observation of CTA 1 with VERITAS}


\author{S. McArthur}
\affiliation{Department of Physics and McDonnell Center for the Space Sciences, Washington University in St. Louis, 1 Brookings Drive, St. Louis, MO 63130, USA}
\author{for the VERITAS Collaboration}
\affiliation{http://veritas.sao.arizona.edu}

\begin{abstract}
CTA~1 (G119.5+10.2) is a composite supernova remnant (SNR) with a shell-type structure in the radio band and a center filled morphology at X-ray energies. \fermi\ has detected a radio-quiet pulsar \psr\  within the radio shell of CTA~1 in a blind search within its first months of operation. Located within an X-ray synchrotron pulsar wind nebula (PWN), the Fermi source is spatially coincident with the \egret\ source \EG. We present the the detection of the system in very-high-energy (VHE) gamma rays by \VERITAS, with a preliminary comparison to other TeV-detected PWNe.

\end{abstract}

\maketitle

\thispagestyle{fancy}


\section{Introduction}
The composite supernova remnant (SNR) CTA~1 (G119.5+10.2) consists of a shell-type structure visible in the radio band with a center filled morphology at X-ray energies. The radio shell, of diameter $\sim 1.8^\circ$~\citep{Sieber1981}, is fainter towards the north-west (NW) of the remnant, possibly due to rapid expansion of the shock into a region of lower density, as supported by HI observations~\citep{Pineault1993}. The distance to CTA~1 is $d=1.4\pm 0.3$ kpc, derived from the associated HI shell~\citep{Pineault1997}. Its age is estimated to be $\sim 1.3\times 10^4$ yr~\citep{Slane2004}. 

Archival X-ray observations of CTA~1 in the 5-10 keV band show non-thermal diffuse emission of low surface brightness in the center of the remnant, likely corresponding to a pulsar wind nebula (PWN) driven by a young pulsar~\citep{Slane1997}. A faint point source, \rxj, is located at the brightest part of the synchrotron emission, and was suggested as a pulsar candidate by Seward~\etal~\citep{Seward1995}. A \chandra\ image of this object provided further evidence of an energetic, rotation-powered pulsar, resolving a central point source, a compact nebula, and a bent jet~\citep{Halpern2004}. 

\subsection{Previous Gamma-Ray Observations}
The earliest association of gamma-ray emission with CTA~1 comes from the detection of the source \EG\ by the \egret\ instrument, with a relatively small 95\% error circle of $28^{\prime}$ \citep{Hartman1999}. Brazier et al.~\citep{Brazier1998} proposed that the gamma-ray emission could originate from a young Geminga-like pulsar, based upon the coincidence with CTA~1, hard spectral index ($\Gamma = 1.58\pm 0.18$ between 70 MeV and 2 GeV), and lack of flux variability. Confirmation of this association came recently when the \fermi\  {\sl Gamma-Ray Space Telescope} discovered the radio-quiet, 316.86 ms gamma-ray pulsar \psr\ in a blind search, using 0.14 years of data~\citep{Abdo2008CTA1}.  Subsequent observations by \XMM\ resulted in the detection of pulsed X-ray emission out of phase with the gamma-ray pulsation~\citep{Caraveo2010},~\citep{Lin2010}. The spin-down power of the pulsar ($\dot{E} = 4.5\times10^{35}$~erg~s$^{-1}$) and characteristic age ($\tau = 1.39\times10^{4}$~yrs) confirmed estimates based on previous observations observations~\citep{Abdo2008CTA1}.

\subsection{Broadband Modeling}
Prompted by the discovery of \psr\ by \fermi, Zhang \etal~\citep{Zhang2009} modeled the pulsed and unpulsed spectral components of the pulsar magnetosphere and PWN.  The pulsed high-energy spectrum was calculated with an outer-gap model and fit to the \egret\ spectrum of Brazier et al.~\citep{Brazier1998}. The unpulsed spectrum of the PWN was calculated with a time-dependent, broken power law injection model with non-thermal emission from synchrotron radiation and inverse Compton scattering of cosmic microwave background (CMB) and ambient infrared (IR) photons.  These calculations predict that the PWN should be detectable in the very-high-energy (VHE) gamma-ray band by \veritas.

\section{CTA 1 imaged by VERITAS}
\subsection{VERITAS observations}
The Very Energetic Radiation Imaging Telescope Array System (\VERITAS) is an array of four 12-meter imaging atmospheric Cherenkov telescapes (IACTs) located at the base camp of the Fred Lawrence Whipple Observatory in southern Arizona.  Each telescope consists of a Davies-Cotton design optical reflector which focuses the Cherenkov light from atmospheric showers onto a camera consisting of 499 photomulitplier tubes and light concentrators with a total FOV of 3$^{\circ}$. \VERITAS\ is able to detect a point source with the strength of 1\% of the Crab Nebula flux at a statistical significance of 5 standard deviation (5$\sigma$) level in approximately 26 hours of observations. \VERITAS\ is sensitive to gamma rays over a wide range of energies (100~GeV to tens of TeV) with an energy resolution of 15-20\%.

\VERITAS\ observed CTA~1 between September 2010 to January 2011 with a total livetime of approximately 26 hours, after selection for good weather conditions and hardware status. Observations were taken in ``wobble'' mode \citep{Fomin1994}, in which the telescope pointing is offset from the source position by some angular distance. An offset distance of 0.7$^\circ$ was used to accommodate the large size of the remnant and the extension of the PWN as seen in X-rays.  Two sets of \emph{a priori} defined gamma-ray/hadronic shower separation cuts, optimized for weak sources of moderate and hard spectra, were applied to the data.  Background was estimated using the ring background model (see, for example, \citep{Berge2007}), with squared angular integration radii of 0.01 deg$^2$ and 0.055 deg$^2$ used for point-source and extended-source searches, respectively. The statistical significance of the excess is calculated using Equation (17) from Li \& Ma~\citep{LiMa1983}.

\subsection{Results}

%%%%%%%%%%%%%
\begin{figure}[t]
\begin{center}
\includegraphics[scale=0.45]{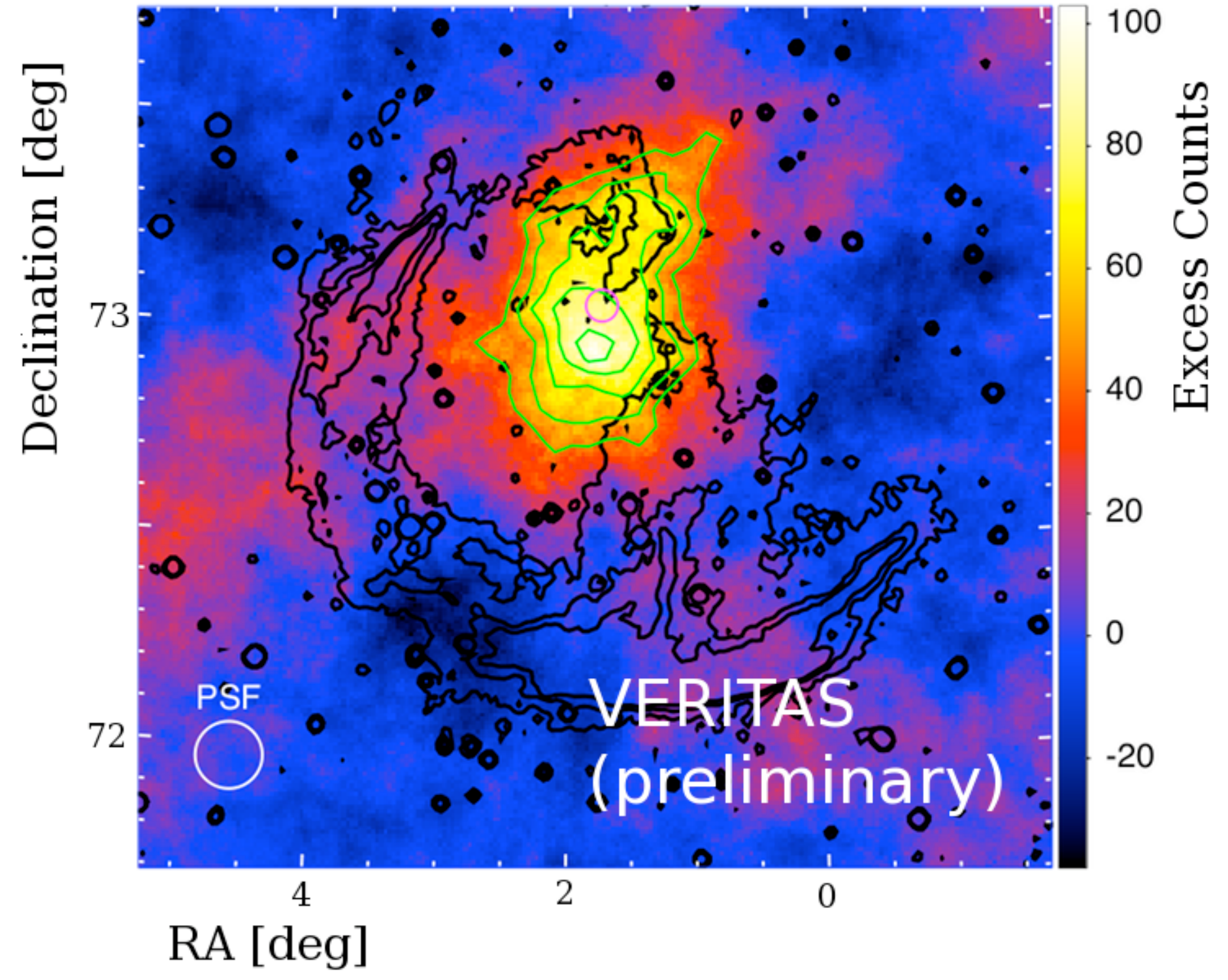}
\caption{\veritas\ excess map of the region around \snr. The color scale indicates excess gamma-ray events in a squared integration radius of 0.055 deg$^2$. The radio contours at 1420 MHz are overlaid in black, showing the SNR shell. The green lines show the \veritas~ significance contours at 3, 4, 5, 6, and 7$\sigma$, respectively. The position of the pulsar is given by the pink circle~\citep{Abdo2008CTA1}.  The circle at the lower left corner shows the size of the \veritas\ PSF (68\% containment).}
\label{fig:CTA1_multi_map}
\end{center}
\end{figure}
%%%%%%%%%%%%%

Figure~\ref{fig:CTA1_multi_map} shows the map of excess events in the region around CTA~1 as measured by \veritas. The hard-spectrum, extended-source analysis produced an excess with a pre-trial significance of $7.3\sigma$, in a blind search region of radius $0.4^\circ$ around the pulsar \psr, within the radio shell of the SNR CTA~1.  Accounting for the sets of cuts and integration radii, and implementing a trails factor for the search region by tiling it with $0.04^\circ$ square bins \citep{Aharonian2006}, we conservatively estimate a post-trials significance of detection of $6.0\sigma$. 

\begin{figure}[t]
\begin{center}
\hspace{-1cm}
\includegraphics[scale=0.45]{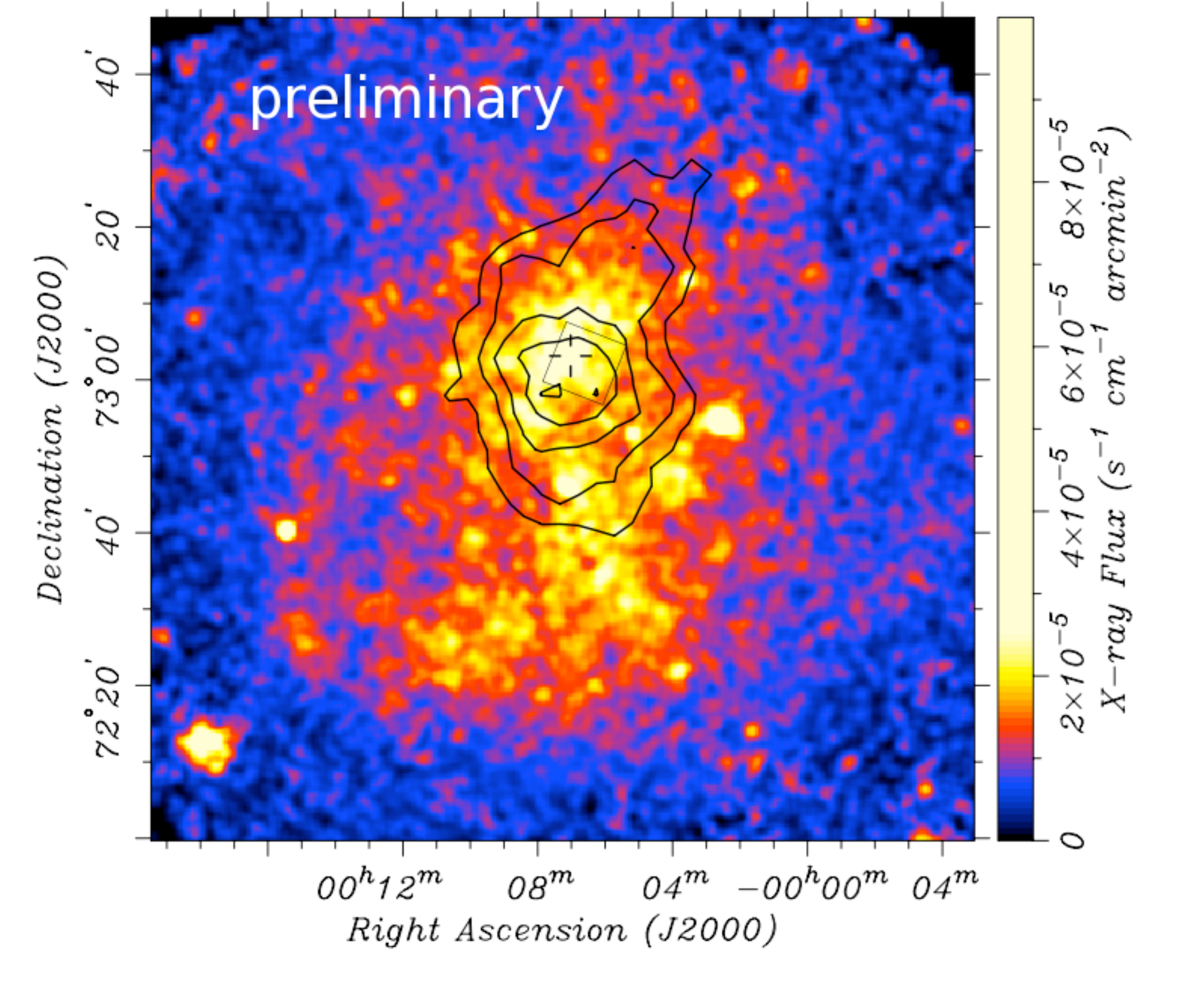}
\caption{\rosat\ X-ray image of the \snr\ shown in equatorial coordinates. The cross marks the location of the X-ray point source \rxj\ and the \fermi\ pulsar. The square shows the field of view for \Chandra. The \veritas\ significance contours for 3 to 7$\sigma$ are shown in black. The \veritas\ excess is seen to line up with location of pulsar.}
\label{fig:cta1rosat}
\end{center}
\end{figure}

The TeV gamma-ray emission region exceeds the point-spread function (PSF; measured from analysis of the Crab Nebula) of \veritas, as seen in Figure~\ref{fig:CTA1_multi_map}. Figure~\ref{fig:cta1rosat} shows the \rosat\ X-ray image of the region around CTA~1, overlaid with the \veritas\ significance contours. The \rosat\ image reveals a center-filled morphology and faint compact source. The \veritas\ excess is roughly centered on the location of \psr, which may be indicative of a young PWN, as opposed to older ``relic" PWNe which have been offset from the pulsar by an interaction with the SNR reverse shock~\citep{Gaensler2006}.  

%%%%%%%%%%%%%
%%%%%%%%%%%%% 

\begin{figure}[t]
\begin{center}
\hspace{-1cm}
\includegraphics[scale=0.35]{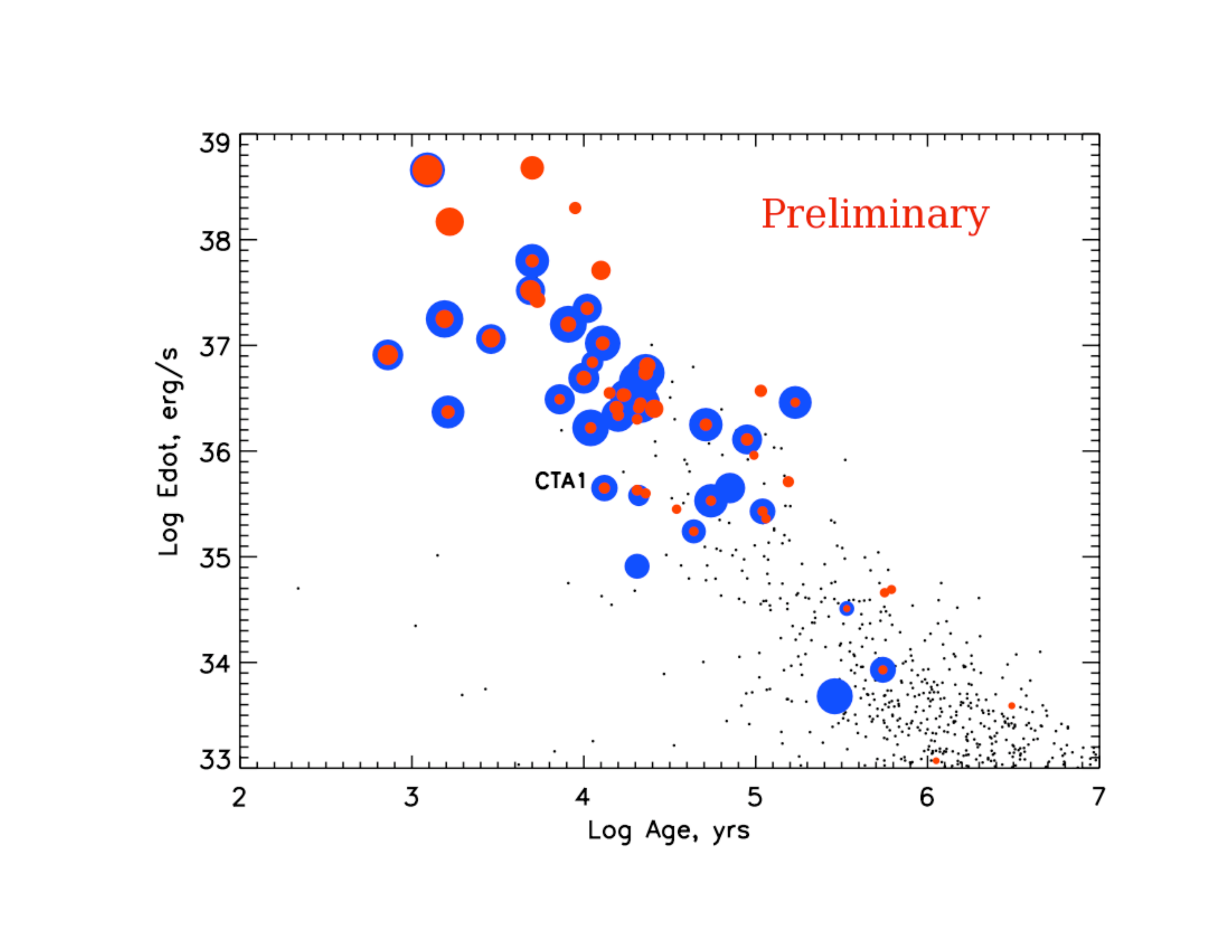}
\caption{Plot of pulsar spin-down luminosity vs age, from Kargaltsev and Pavlov~\citep{Kargaltsev2010}, with CTA 1 point overlaid.  Filled circles: X-ray (red) and TeV (blue) luminosities of PWNe or PWN candidates. Larger circle sizes correspond to higher luminosities in the corresponding waveband. Small black dots denote ATNF catalog pulsars.}
\label{fig:CTA1_EdotVsAge}
\end{center}
\end{figure}

\begin{figure}[t]
\begin{center}
\hspace{-1cm}
\includegraphics[scale=0.35]{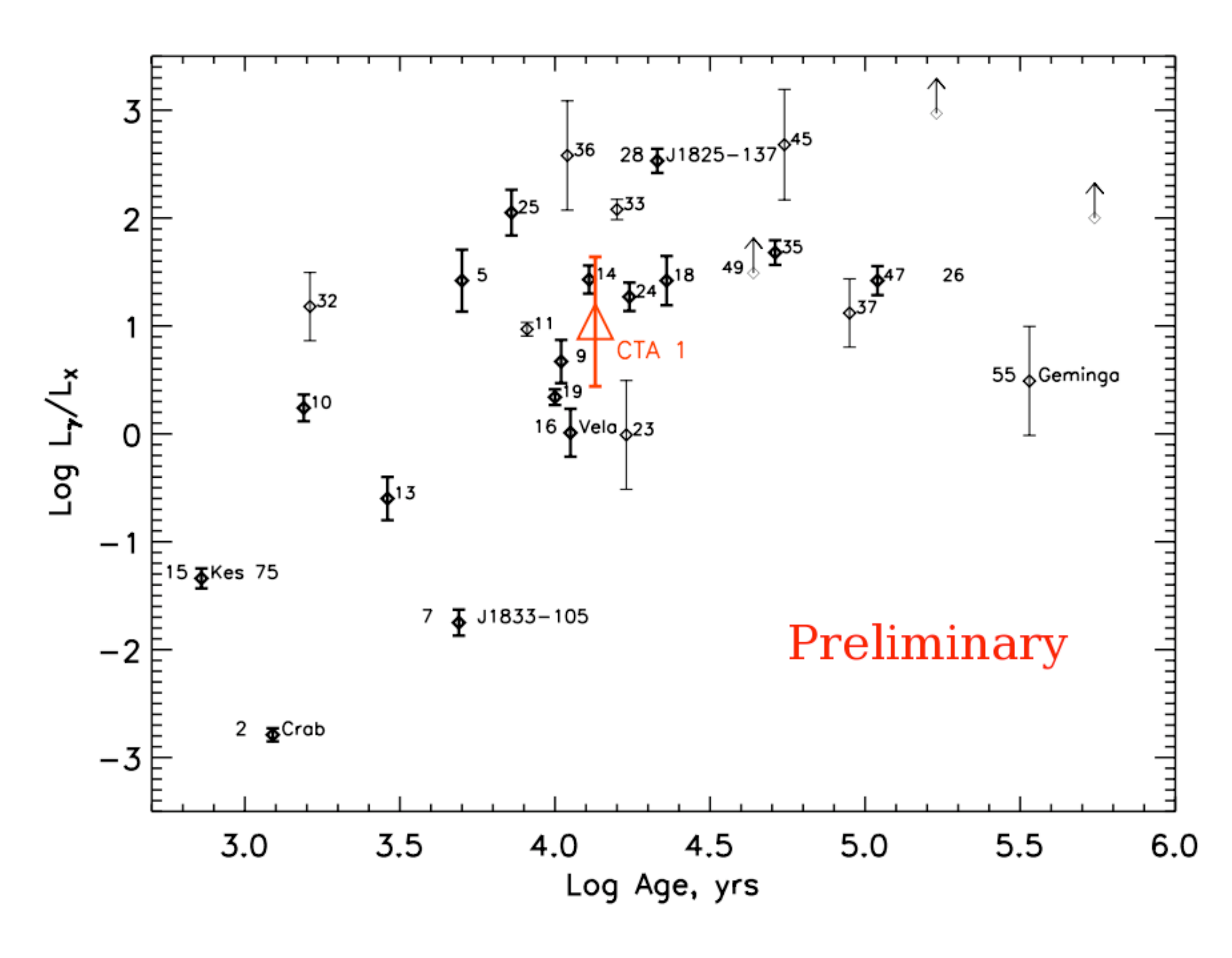}
\caption{Plot of the ratio of TeV to X-ray luminosity vs pulsar spin-down age, from Kargaltsev and Pavlov~\citep{Kargaltsev2010}, with CTA 1 shown by red triangle.}
\label{fig:CTA1_LgVsAge}
\end{center}
\end{figure}

A preliminary spectral analysis gives an integral flux above 1 TeV  of $F_\gamma(>1\mathrm{\ TeV}) \sim 4\%$ of the flux from the Crab Nebula. (Final spectral analysis and flux estimates will be given in a forthcoming paper~\citep{AliuPrep}.) Using the distance of 1.4~kpc, we estimate the luminosity ($L_\gamma = 4\pi d^2 F_\gamma$) to compare with other PWNe and PWNe candidates detected at TeV energies.  Fig.~\ref{fig:CTA1_EdotVsAge} and Fig.~\ref{fig:CTA1_LgVsAge} present the results of these comparisons, following the work of Kargaltsev and Pavlov~\citep{Kargaltsev2010}. Fig.~\ref{fig:CTA1_EdotVsAge} shows the relative luminosities of PWNe in the TeV and X-ray bands, as functions of spin down power and characteristic age. It is seen that TeV PWNe are generally found around younger, more energetic pulsars, although the TeV luminosities do not depend on the pulsar age as strongly as X-ray PWN luminosities do.  Fig.~\ref{fig:CTA1_LgVsAge} shows the distance-independent ratio of TeV gamma-ray luminosity to X-ray luminosity versus the characteristic age. The TeV luminosity of a PWN reflects cumulative pulsar wind properties integrated over a significant fraction of the young pulsar's lifetime while the X-ray luminosity characterizes the freshly injected pulsar wind, which might explain the hint of flattening at larger ages. Again, CTA~1 fits nicely in the middle of the TeV/X-ray PWN population, suggesting that the TeV emission is indeed due to the PWN.

%%%%%%%%%%%%%
%%%%%%%%%%%%%

\section{Summary and Conclusion}
%%%%%%%%%%%%%
%%%%%%%%%%%%%

\veritas\ has detected extended TeV emission within the composite SNR CTA 1 at  a $6\sigma$  post-trials significance level in approximately 26 hours of observation.  The gamma-ray excess lines up with the gamma-ray pulsar \psr, and its X-ray PWN.  Preliminary spectral analysis shows an integral flux above 1 TeV at 4\% of the Crab nebula flux, and the properties of this new TeV source seem consistent with those for the known TeV/X-ray PWN population, lending support to its identification with the PWN of CTA 1.

\bigskip % extra skip inserted
\begin{acknowledgments}
This research is supported by grants from the U.S. Department of Energy Office of Science, the U.S. National Science Foundation and the Smithsonian Institution, by NSERC in Canada, by Science Foundation Ireland (SFI 10/RFP/AST2748) and by STFC in the U.K. We acknowledge the excellent work of the technical support staff at the Fred Lawrence Whipple Observatory and at the collaborating institutions in the construction and operation of the instrument.
\end{acknowledgments}

\bigskip % extra skip inserted

\end{document}